\documentclass[10pt,twocolumn,letterpaper]{article}
\usepackage{iccv}
\usepackage{times}
\usepackage{graphicx}
\usepackage{amsmath}
\usepackage{amssymb}
\usepackage{cases}
\usepackage{setspace}
\usepackage{overpic}
\usepackage{rotating}
\graphicspath{{figures/}}
\usepackage[sort,numbers]{natbib}
\usepackage{subeqnarray}
\usepackage{multirow}
\usepackage{tabularx}
\usepackage{epstopdf}
\usepackage{subfigure}
\usepackage{enumerate}
\usepackage{here}
\usepackage{bm}
\usepackage{color}
\usepackage{xcolor}
\usepackage{colortbl,booktabs}

\usepackage[pagebackref=true,breaklinks=true,letterpaper=true,colorlinks,bookmarks=false]{hyperref}

\iccvfinalcopy 

\def\iccvPaperID{} 

\begin{document}

\title{Designing a Practical Degradation Model for Deep Blind\\ Image Super-Resolution}

\author{Kai Zhang$^1$\qquad\quad Jingyun Liang$^1$\qquad\quad  Luc Van Gool$^{1,2}$\qquad\quad  Radu Timofte$^1$\\
$^1$Computer Vision Lab, ETH Zurich, Switzerland \qquad $^2$KU Leuven, Belgium\\
{\tt\small \{kai.zhang, jinliang, vangool, timofter\}@vision.ee.ethz.ch}\\
\url{https://github.com/cszn/BSRGAN}
}

\maketitle

\begin{abstract}
It is widely acknowledged that single image super-resolution (SISR) methods would not perform well if the assumed degradation model deviates from those in real images. Although several degradation models take additional factors into consideration, such as blur, they are still not effective enough to cover the diverse degradations of real images. To address this issue, this paper proposes to design a more complex but practical degradation model that consists of randomly shuffled blur, downsampling and noise degradations. Specifically, the blur is approximated by two convolutions with isotropic and anisotropic Gaussian kernels; the downsampling is randomly chosen from nearest, bilinear and bicubic interpolations; the noise is synthesized by adding Gaussian noise with different noise levels, adopting JPEG compression with different quality factors, and generating processed camera sensor noise via reverse-forward camera image signal processing (ISP) pipeline model and RAW image noise model. To verify the effectiveness of the new degradation model, we have trained a deep blind ESRGAN super-resolver and then applied it to super-resolve both synthetic and real images with diverse degradations. The experimental results demonstrate that the new degradation model can help to significantly improve the practicability of deep super-resolvers, thus providing a powerful alternative solution for real SISR applications.
\end{abstract}

\section{Introduction}
\label{sec:introduction}

Single image super-resolution (SISR), which aims to reconstruct the natural and sharp detailed high-resolution (HR) counterpart $\mathbf{x}$ from a low-resolution (LR) image $\mathbf{y}$~\cite{timofte2014a+,dong2014learning}, has recently drawn significant attention due to its high practical value. With the advance of deep neural networks (DNNs), there is a dramatic upsurge of using feed-forward DNNs for fast and effective SISR~\cite{zhang2018residual,lim2017enhanced,wang2018esrgan,lai2017deep,hui2019lightweight,liang2021hierarchical}. This paper contributes to this strand.

Whereas SISR methods map an LR image onto an HR counterpart, degradation models define how to map an HR image to an LR one. Two representative degradation models are bicubic degradation~\cite{timofte2017ntire} and traditional degradation~\cite{liu2013bayesian,shocher2018zero}.
The former generates an LR image via bicubic interpolation. The latter can be mathematically modeled by
\begin{equation}\label{eq:sisr_degradation}
  \mathbf{y}\!= \!(\mathbf{x}\otimes \mathbf{k})\!\downarrow_{\bf{s}}\! + \,\mathbf{n}.
\end{equation}
It assumes the LR image is obtained by first convolving the HR image with a Gaussian kernel (or point spread function) $\mathbf{k}$~\cite{efrat2013accurate} to get a blurry image $\mathbf{x}\otimes \mathbf{k}$, followed by a downsampling operation $\downarrow_{\bf{s}}$ with scale factor $\bf{s}$ and an addition of white Gaussian noise $\mathbf{n}$ with standard deviation $\sigma$.
Specifically, the bicubic degradation can be viewed as a special case of traditional degradation as it can be approximated by setting a proper kernel with zero noise~\cite{zhang2020deep,bell2019blind}. The degradation model is generally characterized by several factors such as blur kernel and noise level. Depending on whether these factors are known beforehand or not, DNNs-based SISR methods can be broadly divided into non-blind methods and blind ones.

Early non-blind SISR methods were mainly designed for bicubic degradations~\cite{dong2014learning}. Although significant improvements on the PSNR~\cite{lim2017enhanced,zhang2018residual} and perceptual quality~\cite{wang2018esrgan,ledig2016photo} have been achieved, such methods usually do not perform well on real images.
It is worth noting that this also holds for deep models trained with a generative adversarial loss.
The reason is that blur kernels play a vital role for the success of SISR methods~\cite{efrat2013accurate} and a bicubic kernel is too simple.
To remedy this, some works use a more complex degradation model which involves a blur kernel and additive white Gaussian noise (AWGN) and a non-blind network that takes the blur kernel and noise level as conditional inputs~\cite{zhang2018learning,bell2019blind}. Compared to methods based on bicubic degradation, these tend to be more applicable. Yet, they need an accurate estimation of the kernel and the noise level. Otherwise the performance deteriorates seriously~\cite{efrat2013accurate}. Meanwhile, only a few methods are specially designed for the kernel estimation of SISR~\cite{bell2019blind}. As a further step, some blind methods propose to fuse the kernel estimation into the network design~\cite{gu2019blind,luo2020unfolding}. But such methods still fail to produce visually pleasant results for most real images such as JPEG compressed ones. Along another line of blind SISR work with unpaired LR/HR training data, the kernel and the noise are first extracted from the LR images and then used to synthesize LR images from the HR images for paired training~\cite{ji2020real}. Notably, without kernel estimation, the blind model still has a promising performance. On the other hand, it is difficult to collect accurate blur kernels and noise models from real images.
From the above discussion, we draw two conclusions. Firstly, the degradation model is of vital importance to DNNs-based SISR methods and a more practical degradation model is worth studying.
Secondly, no existing blind SISR models are readily applicable to super-resolve real images suffering from different degradation types.
Hence, we see two main challenges: the first is to design a more practical SISR degradation model for real images, and the second is to learn an effective deep blind model that can work well for most real images. In this paper, we attempt to solve these two challenges.

For the first challenge, we argue that blur, downsampling and noise are the three key factors that contribute to the degradation of real images. Rather than utilizing Gaussian kernel induced blur, bicubic downsampling, and simple noise models, we propose to expand each of these factors to more practical ones.
Specifically, the blur is achieved by two convolutions with an isotropic Gaussian kernel and an anisotropic Gaussian kernel; the downsampling is more general but includes commonly-used downscaling operators such as bilinear and bicubic interpolations; the noise is modeled by AWGN with different noise levels, JPEG compression noise with different quality factors, and processed camera sensor noise by applying reverse-forward camera image signal processing (ISP) pipeline model and RAW image noise model. Furthermore, instead of using the commonly-used blur/downsampling/noise-addition pipeline, we perform randomly shuffled degradations to synthesize LR images.
As a result, our new degradation model involves several more adjustable parameters and aims to cover the degradation space of real images.

For the second challenge, we train a deep model based on the new degradation model in an end-to-end supervised manner.
Given an HR image, we can synthesize different realistic LR images by setting different parameters for the degradation model. As such, an unlimited number of paired LR/HR training data can be generated for training.
Especially noteworthy is that such training data do not suffer from the misalignment issue.
By further taking advantage of the powerful expressiveness and advanced training of DNNs, the deep blind model is expected to produce visually pleasant results for real LR images.

The contributions of this paper are:
\begin{itemize}
  \item[1)] A practical SISR degradation model for real images is designed. It considers more complex degradations for blur, downsampling and noise and, more importantly, involves a degradation shuffle strategy.
  \item[2)] With synthetic training data generated using our degradation model, a blind SISR model is trained. It performs well on real images under diverse degradations.
  \item[3)] To the best of our knowledge, this is the first work to adopt a new hand-designed degradation model for general blind image super-resolution.
  \item[4)] Our work highlights the importance of accurate degradation modeling for practical applications of DNNs-based SISR methods.
\end{itemize}

\section{Related Work}
\label{sec:related_work}
Since this paper focuses on designing a practical degradation model to train a deep blind DNN model, we will next give a brief overview on related degradation models and deep blind SISR methods.

\subsection{Degradation Models}
\label{ssc:degradation_models}

As mentioned in the introduction, existing DNNs-based SISR methods are generally based on bicubic downsampling~\cite{lai2017deep,sajjadi2017enhancenet} and traditional degradations~\cite{zhang2018image,michaeli2013nonparametric,zhang2015revisiting,zhang2019deep,liang2021flow}, or some simple variants~\cite{zhang2017learning,dong2013nonlocally,peleg2014statistical,zhang2018learning,zhang2021plug}. It can be found that existing complex SISR degradation models usually consist of a sequence of blur, downsampling and noise addition. For mathematical convenience, the noise is usually assumed to be AWGN which rarely matches the noise distribution of real images. Indeed, the noise could also stem from camera sensor noise and JPEG compression noise which are usually signal-dependent and non-uniform~\cite{plotz2017benchmarking}.
\textbf{\emph{Regardless of whether the blur is accurately modeled or not, the noise mismatch suffices to cause a performance drop when super-resolvers are applied to real images}}.
In other words, existing degradation models are wanting when it comes to the complexity of real image degradations.
Some works do not consider an explicit degradation model~\cite{yuan2018unsupervised,lugmayr2019unsupervised}. Instead, they use training data to learn the LR-to-HR mapping which only works for the degradations defined by the training images.

\subsection{Deep Blind SISR Methods}
\label{ssc:flexible_SISR}

Significant achievements resulted from the design and training of deep non-blind SISR networks. This said, applying them for blind SISR is a non-trivial issue. It should be noted that blind SISR methods are mainly deployed for real SISR applications. To that end, different research directions have been tried.

The first direction is to initially estimate the degradation parameters for a given LR image, and then apply a non-blind method to obtain the HR result. Bell-Kligler~\etal~\cite{bell2019blind} propose to estimate the blur kernel via an internal-GAN method before applying the non-blind ZSSR~\cite{shocher2018zero} and SRMD~\cite{zhang2018learning} methods. Yet, non-blind SISR methods are usually sensitive to errors in the blur kernel, producing over-sharp or over-smooth results.

To remedy this, a second direction aims to jointly estimate the blur kernel and the HR image. Gu~\etal~\cite{gu2019blind} propose an iterative correction scheme to alternately improve the blur kernel and HR result.
Cornillere~\etal~\cite{cornillere2019blind} propose an optimization procedure for joint blur kernel and HR image estimation by minimizing the error predicted by a trained kernel discriminator. Luo~\etal~\cite{luo2020unfolding} propose a deep alternating network that consists of a kernel estimator module and an HR image restorer module.
While promising, these methods do not fully take noise into consideration and thus tend to suffer from inaccurate kernel estimation for noisy real images. As a matter of fact, the presence of noise would aggravate the ill-posedness, especially when the noise type is unknown and complex, and the noise level is high.

A third direction is to learn a supervised model with captured real LR/HR pairs. Cai~\etal~\cite{cai2019toward} and Wei~\etal~\cite{wei2020aim} separately established a SISR dataset with paired LR/HR camera images. Collecting abundant well-aligned training data is cumbersome however, and the learned models are constrained to the LR domain defined by the captured LR images.

Considering the fact that real LR images rarely come with the ground-truth HR, the fourth direction aims at learning with unpaired training data~\cite{wang2021unsupervised}.
Yuan~\etal~\cite{yuan2018unsupervised} propose a cycle-in-cycle framework to first map the noisy and blurry LR input to a clean one and then super-resolve the intermediate LR image via a pre-trained model.
Lugmayr~\etal~\cite{lugmayr2019unsupervised} propose to learn a deep degradation mapping by employing a cycle consistency loss and then generate LR/HR pairs for supervised training.
Following a similar framework, Ji~\etal~\cite{ji2020real} propose to estimate various blur kernels and extract different noise maps from LR images and then apply the traditional degradation model to synthesize different LR images. Notably, \cite{ji2020real} was the winner of the NTIRE 2020 real-world super-resolution challenge~\cite{lugmayr2020ntire}, which demonstrates the importance of accurate degradation modeling. Although applying this method to training data corrupted by a more complex degradation seems to be straightforward, it would also reduce the accuracy of blur kernel and noise estimation which in turn results in unreliable synthetic LR images.

As discussed above, existing deep blind SISR methods are mostly trained on ideal degradation settings or specific degradation spaces defined by the LR training data.
As a result, there is still a mismatch between the assumed degradation model and the real image degradation model.
Furthermore, to the best of our knowledge, \textbf{\emph{no existing deep blind SISR model can be readily applied for general real image super-resolution}}.
Therefore, it is worthwhile to design a practical degradation model to train deep blind SISR models for real applications.
Note that, although denoising and deblurring are related to noisy and blurry image super-resolution, most super-resolution methods tackle the blur, noise and super-resolution in a unified rather than a cascaded framework (see, \eg, ~\cite{liu2013bayesian,efrat2013accurate,dong2013nonlocally,zhang2017learning,riegler2015conditioned,shocher2018zero,zhang2018learning,zhang2020deep,ji2020real,yuan2018unsupervised,lugmayr2019unsupervised,lugmayr2020ntire}).

\section{A Practical Degradation Model}
\label{sec:method}
Before providing our new practical SISR degradation model, it is useful to mention the following facts on the bicubic and traditional degradation models:
\begin{enumerate}
\item According to the traditional degradation model, there are three key factors, \ie, blur, downsampling and noise, that affect the degradations of real images.
\item Since both LR and HR images could be noisy and blurry, it is not necessary to adopt the blur/downsampling/noise-addition pipeline as in the traditional degradation model to generate LR images.
\item The blur kernel space of the traditional degradation model should vary across scales, making it in practice tricky to determine for very large scale factors.
\item While the bicubic degradation is rarely suitable for real LR images, it can be used for data augmentation and is indeed a good choice for clean and sharp image super-resolution.
\end{enumerate}

Inspired by the first fact, a direct way to improve the practicability of degradation models is to make  the degradation space of the three key factors as large and realistic as possible.
Based on the second fact, we then further expand the degradation space by adopting a random shuffle strategy for the three key factors. Like that, an LR image could also be a noisy, downsampled and blurred version of the HR image.
To tackle the third fact,
one may take advantage of the analytical calculation of the kernel for a large scale factor from a small one.
Alternatively, according to the fourth fact, for a large scale factor, one can apply a bicubic (or bilinear) downscaling before the degradation with scale factor 2.
Without loss of generality, this paper focuses on designing the degradation model for the widely-used scale factors 2 and 4.

In the following, we will detail the degradation model for the following aspects: blur, downsampling, noise, and random shuffle strategy.

\subsection{Blur}\label{section:blur}
Blur is a common image degradation. We propose to model the blur from \textbf{\emph{both the HR space and LR space}}. On the one hand, in the traditional SISR degradation model~\cite{liu2013bayesian,shocher2018zero}, the HR image is first blurred by a convolution with a blur kernel. This HR blur actually aims to prevent aliasing and preserve more spatial information after the subsequent downsampling. On the other hand, the real LR image could be blurry and thus it is a feasible way to model such blur in the LR space. By further considering that Gaussian kernels suffice for the SISR task, we perform \textbf{\emph{two Gaussian blur operations}}, \ie, $\mathbf{B}_{\text{iso}}$ with isotropic Gaussian kernels and $\mathbf{B}_{\text{aniso}}$ with anisotropic Gaussian kernels~\cite{zhang2018learning,bell2019blind,riegler2015conditioned}. Note that the HR image or LR image could be blurred by two blur operations (see Sec.~\ref{ssc:random_shuffle} for more details). By doing so, the degradation space of blur can be greatly expanded.

For the blur kernel setting, the size is uniformly sampled from \{$7\times7$, $9\times9$, $\cdots$, $21\times21$\},
the isotropic Gaussian kernel samples the kernel width uniformly from $[0.1, 2.4]$ and $[0.1, 2.8]$ for scale factors 2 and 4, respectively, while the anisotropic Gaussian kernel samples the rotation angle uniformly from $[0, \pi]$ and the length of each axis for scale factors 2 and 4 uniformly from $[0.5, 6]$ and $[0.5, 8]$, respectively.
Reflection padding is adopted to ensure the spatial size of the blurred output stays the same. Since the isotropic Gaussian kernel with width $0.1$ corresponds to delta (identity) kernel, we can always apply the two blur operations.

\subsection{Downsampling}
\label{section:downsampling}

In order to downsample the HR image, perhaps the most direct way is nearest neighbor interpolation. Yet, the resulting LR image will have a misalignment of $0.5$$\times$$(\bf{s}-\text{1})$ pixels towards the upper-left corner~\cite{zhang2020deep}. As remedy, we shift a centered $21\times21$ isotropic Gaussian kernel by $0.5$$\times$$(\bf{s}-\text{1})$ pixels via a 2D linear grid interpolation method~\cite{liu2013bayesian}, and apply it for convolution before the nearest neighbour downsampling. The Gaussian kernel width is randomly chosen from $[0.1, 0.6\times \bf{s}]$.
We denote such a downsampling as $\mathbf{D}^{\bf{s}}_{\text{nearest}}$.
In addition, we also adopt the bicubic and bilinear downsampling methods, denoted by $\mathbf{D}^{\bf{s}}_{\text{bilinear}}$ and $\mathbf{D}^{\bf{s}}_{\text{bicubic}}$, respectively.
Furthermore, a down-up-sampling method $\mathbf{D}^{\bf{s}}_{\text{down-up}}(=\mathbf{D}^{\bf{s}/{a}}_{\text{down}}\mathbf{D}^{\bf{a}}_{\text{up}})$ which first downsamples the image with a scale factor $\bf{s}/{a}$ and then upscales with a scale factor $\bf{a}$ is also adopted. Here the interpolation methods are randomly chosen from bilinear and bicubic interpolations, and $\bf{a}$ is sampled from $[1/2, \bf{s}]$. Clearly, the above four downsampling methods have a blurring step in the HR space, while $\mathbf{D}^{\bf{s}}_{\text{down-up}}$ can introduce upscaling-induced blur in the LR space when $\bf{a}$ is smaller than $1$.
We do not include such kinds of blur in Sec.~\ref{section:blur} since they are coupled in the downsampling process. We uniformly sample these four downsampling to downscale the HR image.

\subsection{Noise}
\label{section:noise}

\begin{figure*}[!tbp]
\begin{center}
\begin{overpic}[width=0.99\textwidth]{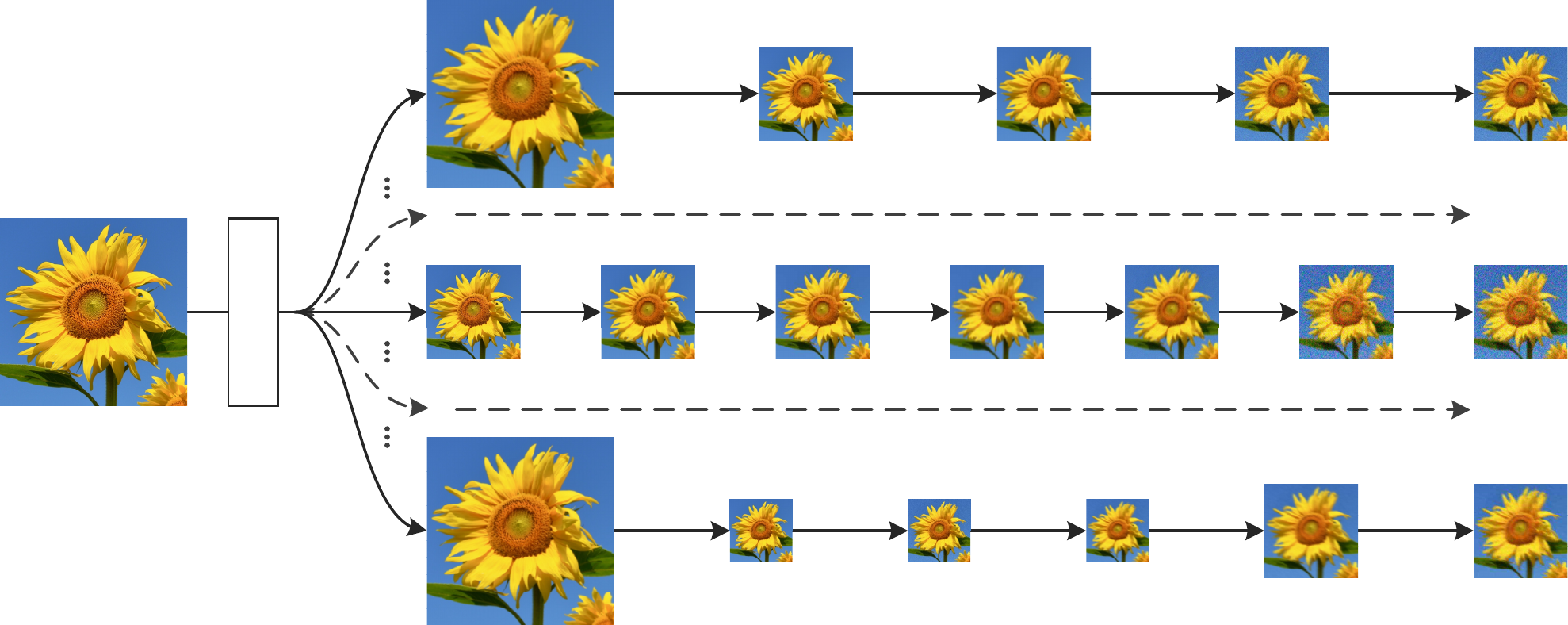}
\put(4.6,12){\color{black}{\footnotesize HR}}
\put(96,13.1){\color{black}{\footnotesize LR}}
\put(96,25.6){\color{black}{\footnotesize LR}}

\put(19.1,28.5){\color{black}{\scriptsize $\mathbf{B}_{\text{aniso}}$}}
\put(41.4,34.5){\color{black}{\scriptsize $\mathbf{D}^2_{\text{bilinear}}$}}
\put(56.8,34.5){\color{black}{\scriptsize $\mathbf{B}_{\text{iso}}$}}
\put(72.6,34.5){\color{black}{\scriptsize $\mathbf{N}_{\text{G}}$}}
\put(86.8,34.5){\color{black}{\scriptsize $\mathbf{N}_{\text{JPEG}}$}}

\put(21.6,20.7){\color{black}{\scriptsize $\mathbf{D}^2_{\text{bicubic}}$}}
\put(34,20.7){\color{black}{\scriptsize $\mathbf{B}_{\text{iso}}$}}
\put(44.5,20.7){\color{black}{\scriptsize $\mathbf{N}_{\text{JPEG}}$}}
\put(55.8,20.7){\color{black}{\scriptsize $\mathbf{B}_{\text{aniso}}$}}
\put(68,20.7){\color{black}{\scriptsize $\mathbf{N}_{\text{G}}$}}
\put(78.9,20.7){\color{black}{\scriptsize $\mathbf{N}_{\text{S}}$}}

\put(19,10.6){\color{black}{\scriptsize $\mathbf{B}_{\text{iso}}$}}

\put(40.0,6.8){\color{black}{\scriptsize $\mathbf{D}^3_{\text{down}}$}}
\put(52.8,6.8){\color{black}{\scriptsize $\mathbf{N}_{\text{G}}$}}
\put(63.6,6.8){\color{black}{\scriptsize $\mathbf{B}_{\text{aniso}}$}}
\put(74.2,6.8){\color{black}{\scriptsize $\mathbf{D}^{2/3}_{\text{up}}$}}
\put(87.8,6.8){\color{black}{\scriptsize $\mathbf{N}_{\text{JPEG}}$}}

\put(89.1,20.6){\color{black}{\scriptsize $\mathbf{N}_{\text{JPEG}}$}}

\begin{turn}{270}
\put(-25.7,15.8){\color{black}{\scriptsize Degradation Shuffle}}

\end{turn}
\end{overpic}
\end{center}\vspace{-0.1cm}
\caption{Schematic illustration of the proposed degradation model for scale factor 2. For an HR image, the randomly shuffled degradation sequences $\{\mathbf{B}_{\texttt{iso}}, \mathbf{B}_{\texttt{aniso}}, \mathbf{D}^2, \mathbf{N}_{\text{G}}, \mathbf{N}_{\text{JPEG}}, \mathbf{N}_{\text{S}}\}$ are first performed, then a JPEG compression degradation $\mathbf{N}_{\text{JPEG}}$ is applied to save the LR image into JPEG format. The downscaling operation with scale factor 2, \ie, $\mathbf{D}^2$, is uniformly chosen from $\{\mathbf{D}^2_{\text{nearest}}, \mathbf{D}^2_{\text{bilinear}}, \mathbf{D}^2_{\text{bicubic}}, \mathbf{D}^{2}_{\text{down-up}}\}$. }
\label{fig:degradationmodel}
\end{figure*}
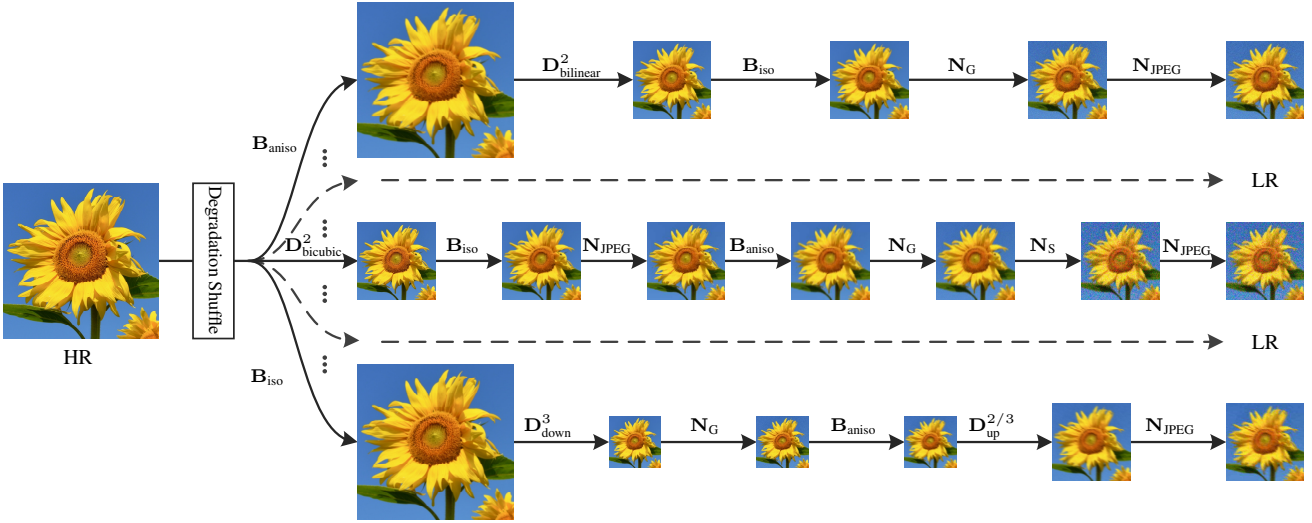

Noise is ubiquitous in real images as it can be caused by different sources. Apart from the widely-used Gaussian noise, our new degradation model also considers JPEG compression noise and camera sensor noise. We next detail the three noise types.

\begin{spacing}{1.5}
\end{spacing}
\noindent
\textbf{Gaussian noise $\mathbf{N}_{\text{G}}$.}
The Gaussian noise assumption is the most conservative choice when there is no information about the noise~\cite{park2013gaussian}.
To synthesize Gaussian noise, the three-dimensional (3D) zero-mean Gaussian noise model  $\mathcal{N}(\mathbf{0},\mathbf{\Sigma})$~\cite{nam2016holistic} with covariance matrix $\mathbf{\Sigma}$ is adopted.
Such noise model has two special cases: when $\mathbf{\Sigma}=\sigma^2\bf{I}$, where $\bf{I}$ is the identity matrix, it turns into the widely-used channel-independent additive white Gaussian noise (AWGN) model; when $\mathbf{\Sigma}=\sigma^2\bf{1}$, where $\bf{1}$ is a $3\times3$ matrix with all elements equal to one, it turns into the widely-used gray-scale AWGN model.
In our new degradation model, we always add Gaussian noise for data synthesis. In particular, the probabilities of applying the general case and two special cases are set to 0.2, 0.4, 0.4, respectively. As for $\sigma$, it is uniformly sampled from $\{1/255,2/255, \cdots, 25/255\}$.

\begin{spacing}{1.5}
\end{spacing}
\noindent
\textbf{JPEG compression noise $\mathbf{N}_{\text{JPEG}}$.}
JPEG is the most widely-used image compression standard for bandwidth and storage reduction. Yet, it introduces annoying $8\times8$ blocking artifacts/noise, especially for the case of high compression.
The degree of compression is determined by the quality factor which is an integer in the range $[0, 100]$. The quality factor 0 means lower quality and higher compression, and vice versa. If the quality factor is larger than 90, no obvious artifacts are introduced.
In our new degradation model, the JPEG quality
factor is uniformly chosen from $[30, 95]$. Since JPEG is the most popular digital image format, we apply \textbf{\emph{two}} JPEG compression steps with possibilities 0.75 and 1, respectively. In particular, the latter one is used as the final degradation step.

\begin{spacing}{1.5}
\end{spacing}
\noindent
\textbf{Processed camera sensor noise $\mathbf{N}_{\text{S}}$.}
In modern digital cameras, the output image is obtained by passing the raw sensor data through the image signal processing (ISP) pipeline. In practice, if the ISP pipeline does not perform a denoising step, the processed sensor noise would deteriorate the output image by introducing non-Gaussian noise~\cite{plotz2017benchmarking}. To synthesize such kind of noise, we first get the raw image from an RGB image via the reverse ISP pipeline, and then reconstruct the noisy RGB image via the forward pipeline after adding noise to the synthetic raw image. The raw image noise model is borrowed from~\cite{brooks2019unprocessing}. According to the Adobe Digital Negative (DNG) Specification~\cite{adobe2019}, our forward ISP pipeline consists of demosaicing, exposure compensation, white balance, camera to XYZ (D50) color space conversion, XYZ (D50) to linear RGB color space conversion, tone mapping and gamma correction. For demosaicing,
the method in~\cite{malvar2004high} which is the same as matlab's \texttt{demosaic} function, is adopted. For exposure compensation, the global scaling is chosen from $[2^{-0.1}, 2^{0.3}]$. For the white balance, the red gain and blur gain are uniformly chosen from $[1.2, 2.4]$. For camera to XYZ (D50) color space conversion, the $3\times3$ color correction matrix is a random weighted combination of \texttt{ForwardMatrix1} and \texttt{ForwardMatrix2} from the metadata of raw image files.
For the tone mapping, we manually select the best fitted tone curve from~\cite{grossberg2003space} for each camera based on paired raw image files and the RGB output. We use five digital cameras, including the Canon EOS 5D Mark III and IV cameras, Huawei P20, P30 and Honor V8 cameras, to establish our ISP pipeline pool. Note that the tone curve and forward color correction matrix do not necessarily come from the same camera. Since tone mapping is not reversible and would result in color shift issue, one should apply the reverse-forward tone mapping for the HR image.
We apply this noise synthesis step with a probability of 0.25.

\subsection{Random Shuffle}
\label{ssc:random_shuffle}
Though simple and mathematically convenient, the traditional degradation model can hardly cover the degradation space of real LR images. On the one hand, the real LR image could also be a noisy, blurry, downsampled, and JPEG compressed version of the HR image. On the other hand, the degradation model which assumes the LR image is a bicubicly downsampled, blurry and noisy version of the HR image can also be used for SISR~\cite{gu2019blind,zhang2019deep}.
Hence, an LR image can be degraded by blur, downsampling, and noise with different orders. We thus propose a random shuffle strategy for the new degradation model. Specifically, the degradation sequence
$\{\mathbf{B}_{\texttt{iso}}, \mathbf{B}_{\texttt{aniso}}, \mathbf{D}^{\bf{s}}, \mathbf{N}_{\text{G}}, \mathbf{N}_{\text{JPEG}}, \mathbf{N}_{\text{S}}\}$ is randomly shuffled, here $\mathbf{D}^{\bf{s}}$ represents the downsampling operation with scale factor $\bf{s}$ which is randomly chosen from $\{\mathbf{D}^{\bf{s}}_{\text{nearest}}, \mathbf{D}^{\bf{s}}_{\text{bilinear}}, \mathbf{D}^{\bf{s}}_{\text{bicubic}}, \mathbf{D}^{\bf{s}}_{\text{down-up}}\}$. In particular, the sequence of $\mathbf{D}^{\bf{s}/{a}}_{\text{down}}$ and $\mathbf{D}^{\bf{a}}_{\text{up}}$ for $\mathbf{D}^{\bf{s}}_{\text{down-up}}$ can insert other degradations.
Note that a similar idea of random shuffle strategy was proposed in~\cite{cubuk2020randaugment}, however, it is designed for image classification and object detection and could be instead used to augment HR images.

With the random shuffle strategy, the degradation space can be expanded substantially.
Firstly, other degradation models, such as bicubic and traditional degradation models, and the ones proposed in~\cite{gu2019blind,zhang2019deep}, are special cases of ours. Secondly, the blur degradation space is enlarged by different arrangements of the two blur operations and one of the four downsampling methods.
Thirdly, the noise characteristics could be changed by the blur and downsampling, thus expanding the degradation space. For example, the downsampling can reduce the noise strength and make the noise (\eg, processed camera sensor noise and JPEG compression noise) less signal-dependent, whereas $\mathbf{D}^{\bf{a}}_{\text{up}}$ ($\bf{a}<\text{1}$) can make the signal-independent Gaussian noise to be signal-dependent. Such kinds of noise could exist in real images.

Fig.~\ref{fig:degradationmodel} illustrates the proposed degradation model. For an HR image, we can generate different LR images with a wide range of degradations by shuffling the degradation operations and setting different degradation parameters. As mentioned in Sec.~\ref{sec:method}, for scale factor 4, we additionally apply a bilinear or bicubic downscaling before the degradation for scale factor 2 with a probability of 0.25.

\section{Discussion}
It is necessary to add discussion to further understand the proposed new degradation model.
\textbf{\emph{Firstly}}, the degradation model is mainly designed to synthesize degraded LR images.
Its most direct application is to train a deep blind super-resolver with paired LR/HR images.
In particular, the degradation model can be performed on a large dataset of HR images to produce unlimited perfectly aligned training images, which typically do not suffer from the limited data issue of laboriously collected paired data and the misalignment issue of unpaired training data.
\textbf{\emph{Secondly}}, the degradation model tends to be unsuited to model a degraded LR image as it involves too many degradation parameters and also adopts a random shuffle strategy.
\textbf{\emph{Thirdly}}, the degradation model can produce some degradation cases that rarely happen in real-world scenarios, while this can still be expected to improve the generalization ability of the trained deep blind super-resolver.
\textbf{\emph{Fourthly}}, a DNN with large capacity has the ability to handle different degradations via a single model (see, \eg,~\cite{zhang2017beyond}). It is worth noting that even when the super-resolver reduces the performance for unrealistic bicubic downsampling, it is still a preferred choice for real SISR.
\textbf{\emph{Fifthly}}, one can conveniently modify the degradation model by changing the degradation parameter settings and adding more reasonable degradation types (\eg,
speckle noise and unaligned double JPEG compression~\cite{jiang2021towards}) to improve the practicability for certain applications.

\section{Deep Blind SISR Model Training}\label{sec:experiments}

The novelty of this paper lies in the new degradation model and the possibility of existing network structures such as ESRGAN~\cite{wang2018esrgan} to be borrowed to train a deep blind model.
For the sake of showing the advantage of the proposed degradation model, we adopt the widely-used ESRGAN network and train it with the synthetic LR/HR paired images produced by the new degradation model.
Following ESRGAN, we first train a PSNR-oriented BSRNet model and then train the perceptual quality-oriented BSRGAN model.
Since the PSNR-oriented BSRNet model tends to produce oversmoothed results due to the pixel-wise average problem~\cite{ledig2016photo}, the perceptual quality-oriented model is preferred for real applications~\cite{blau2018perception}.
Thus, unless otherwise specified, we focus more on the BSRGAN model.

Compared to ESRGAN, BSRGAN is modified in several ways. First, we use a slightly different HR image dataset which includes DIV2K~\cite{agustsson2017ntire}, Flick2K ~\cite{timofte2017ntire,lim2017enhanced},
WED~\cite{ma2016gmad} and 2,000 face images from FFHQ~\cite{karras2019style}
to capture the image prior. The reason is that
the goal of BSRGAN is to solve the problem of general-purpose blind image super-resolution, and apart from the degradation prior, an image prior could also contribute to the success of a super-resolver. We also remove the blurry images based on the variance of the Laplacian of an image.
Secondly, BSRGAN uses a larger LR patch size of $72\times72$. The reason is that our degradation model can produce severely degraded LR images and a larger patch can enable deep models to capture more information for better restoration. Thirdly, we train the BSRGAN by minimizing a weighted combination of L1 loss, VGG perceptual loss and spectral norm-based least square PatchGAN loss~\cite{isola2017image} with weights $1$, $1$ and $0.1$, respectively. In particular, the VGG perceptual loss is operated on the fourth convolution before the fourth rather than the fifth maxpooling layer of the pre-trained 19-layer VGG model as it is more stable to prevent color shift issues.
We train BSRGAN with Adam, using a fixed learning rate of $1\times10^{-5}$ and a batch size of 48.

\begin{figure}[!bp]\small
\hspace{-0.26cm}
\begin{tabular}{c@{\extracolsep{0em}}@{\extracolsep{0.15em}}c}
       \includegraphics[width=0.235\textwidth]{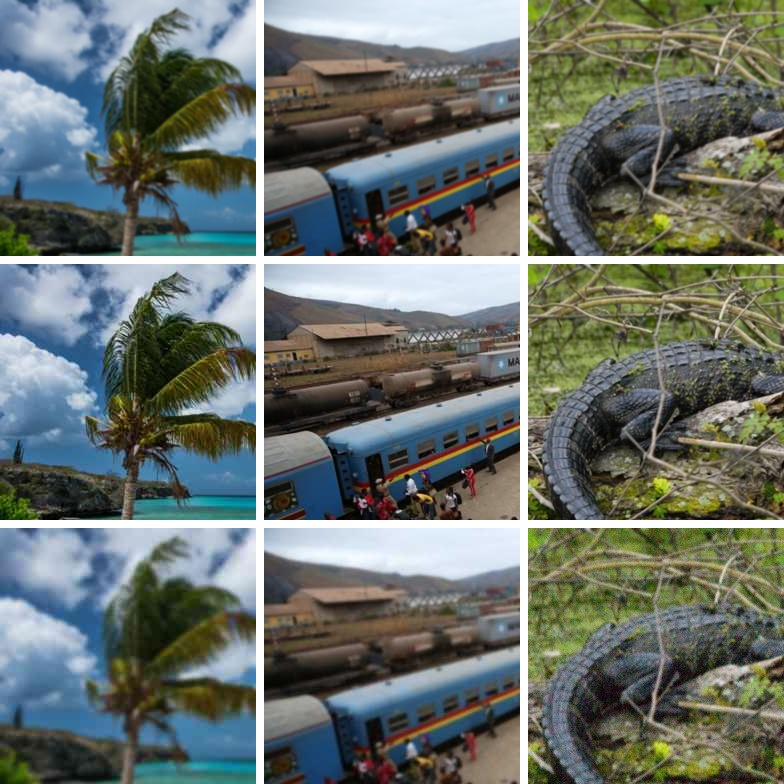}~
		&\includegraphics[width=0.235\textwidth]{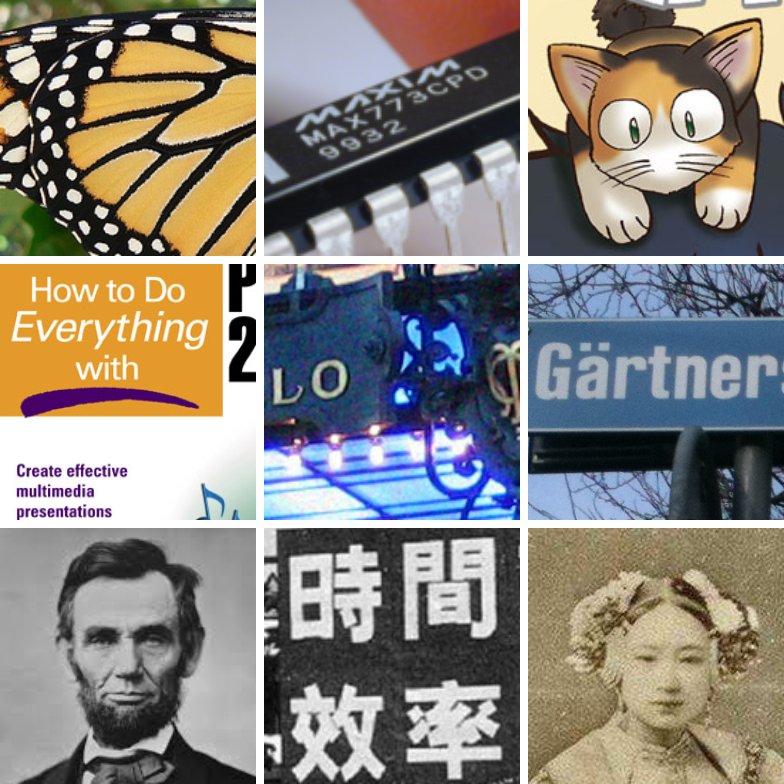}\\
(a) Examples from DIV2K4D  & (b) Examples from RealSRSet \\
	\end{tabular}
    \vspace{0.1cm}
	\caption{Some example images from the DIV2K4D and RealSRSet datasets. From top to bottom of (a), we show example images generated by the degradation types II, III and IV.}
	\label{fig:dataset}
\end{figure}

\section{Experimental Results}

\subsection{Testing Datasets}
Existing blind SISR methods are generally evaluated on specifically designed synthetic data and only very few real images. For example,
IKC~\cite{gu2019blind} is evaluated on the blurred, bicubicly downsampled synthetic LR images and two real images;
KernelGAN~\cite{bell2019blind} is evaluated on the synthetic DIV2KRK dataset and two real images.
As a result, to the best of our knowledge, a real LR image dataset with diverse blur and noise degradations is still lacking.

\begin{table*}[!htbp]\footnotesize
\caption{The PSNR and LPIPS results of different methods on the DIV2K4D dataset. The best and second best results are highlighted in red and blue, respectively. The PSNR results are calculated on Y channel of YCbCr space.}
\vspace{-0.2cm}
\center
\begin{tabular}{p{1.5cm}<{\centering}|p{1.0cm}<{\centering}||p{1.0cm}<{\centering}|p{0.9cm}<{\centering}|p{1.2cm}<{\centering}|p{1.2cm}<{\centering}|p{1.1cm}<{\centering}|p{1.1cm}<{\centering}|p{1.1cm}<{\centering}||p{1.2cm}<{\centering}|p{1.2cm}<{\centering}}
\hline
Degradation   & \multirow{2}{*}{Metric} & \multirow{2}{*}{RRDB}& \multirow{2}{*}{IKC}   & \multirow{2}{*}{ESRGAN}  & FSSR  & FSSR & RealSR  &  RealSR & \textbf{BSRNet} & \textbf{BSRGAN}  \\
 Type &  &  &  &   & -DPED  & -JPEG  & -DPED  & -JPEG & \textbf{(Ours)} & \textbf{(Ours)}  \\ \hline\hline

Type I  & PSNR & \textcolor{red}{30.89} &  \textcolor{blue}{29.95}  & 28.16  & 24.55  & 22.71 & 21.72  & 27.35 & {{29.07}} & 27.30  \\
   (Bicubic)                       & LPIPS & 0.254 & 0.263 & \textcolor{red}{0.115}  & 0.240 & 0.364  & 0.312  &  \textcolor{blue}{0.213}& 0.331   & 0.236  \\ \hline

\multirow{2}{*}{Type II}   & PSNR & 25.66 & \textcolor{blue}{27.35}  & 25.56  & 25.81  & 25.33 & 26.29  & 25.36 & \textcolor{red}{27.76} & 26.26  \\
                          & LPIPS & 0.542 & 0.392 & 0.526  & 0.460 & 0.399  & \textcolor{red}{0.263}  & 0.479& 0.397   & \textcolor{blue}{{0.284}}  \\ \hline

\multirow{2}{*}{Type III}   & PSNR & 26.70 & \textcolor{blue}{26.72}  & 26.21  & 25.83  & 23.25  & 22.82  & 26.72  & \textcolor{red}{27.59}& 26.28  \\
                          & LPIPS & 0.517 & 0.504  & 0.436  & 0.392  & \textcolor{blue}{{0.376}}  & 0.379  & 0.360  & 0.419& \textcolor{red}{0.284}  \\ \hline

\multirow{2}{*}{Type IV}   & PSNR & 24.03 & 24.01 & 23.68  & 23.62  & 22.40 & 22.97  & 23.85 & \textcolor{red}{25.67}  & \textcolor{blue}{24.58}  \\
                          & LPIPS & 0.659 & 0.641  & 0.599  &  0.589 & 0.597  &  0.528 & 0.589 & \textcolor{blue}{0.506} & \textcolor{red}{0.361}  \\ \hline

\end{tabular}
\label{table:div2k4d}
\end{table*}

\begin{figure*}[!htbp]\small
\hspace{-0.20cm}
\begin{tabular}{c@{\extracolsep{0em}}c@{\extracolsep{0.04em}}c@{\extracolsep{0.04em}}c@{\extracolsep{0.04em}}c@{\extracolsep{0.04em}}@{\extracolsep{0.04em}}c@{\extracolsep{0.04em}}c}
        \includegraphics[width=0.16\textwidth]{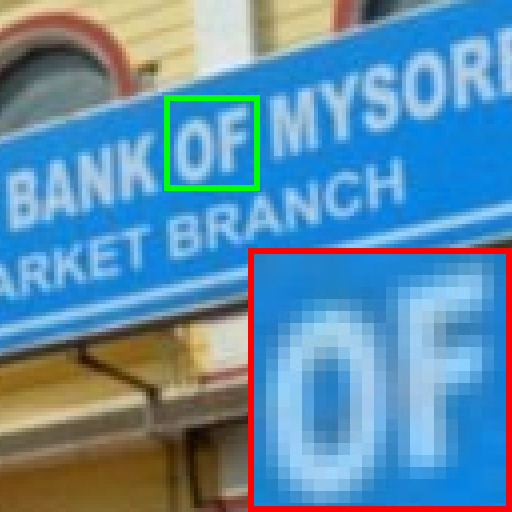}~
		&\includegraphics[width=0.16\textwidth]{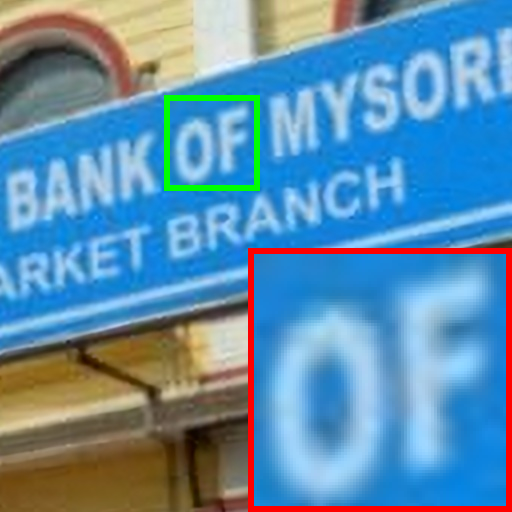}~
		&\includegraphics[width=0.16\textwidth]{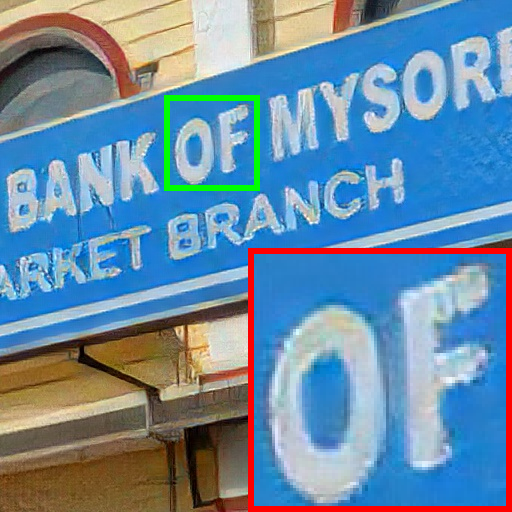}~
        &\includegraphics[width=0.16\textwidth]{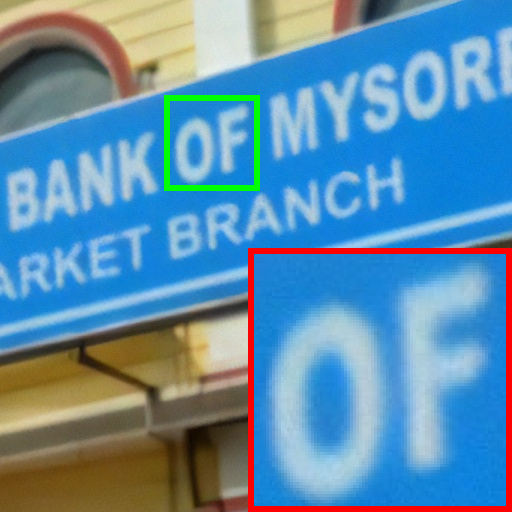}~
		&\includegraphics[width=0.16\textwidth]{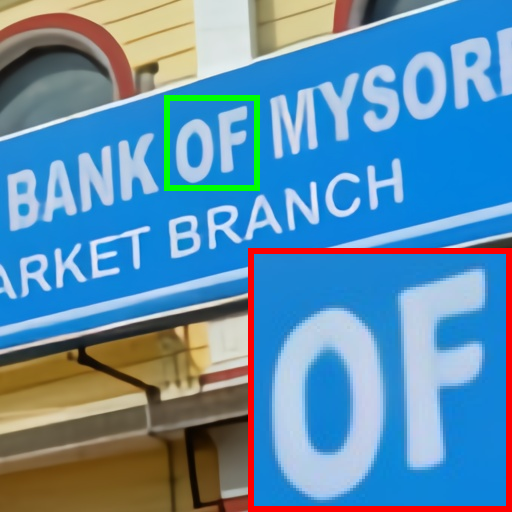}~
		&\includegraphics[width=0.16\textwidth]{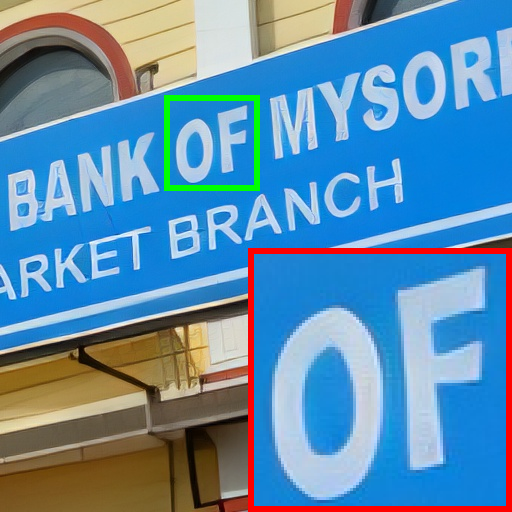}\\
\footnotesize PSNR$\uparrow$/LPIPS$\downarrow$ &23.51/0.601 & 23.21/\textcolor{blue}{0.353}  & 23.46/0.504 & \textcolor{red}{25.48}/\textcolor{blue}{0.353} & \textcolor{blue}{24.65}/\textcolor{red}{0.233} \\
(a) LR ($\times$4) & (b) IKC~\cite{gu2019blind}  & (c) FSSR-JPEG~\cite{fritsche2019frequency} & (d) RealSR-JPEG~\cite{ji2020real} & (e) \textbf{BSRNet (Ours)} &  (f) \textbf{BSRGAN (Ours)} \\
	\end{tabular}
    \vspace{0.001cm}
	\caption{Results of different methods on super-resolving an LR image from the DIV2K4D dataset with scale factor 4. The testing image is synthesized by our proposed degradation (\ie, degradation type IV).}
	\vspace{-0.2cm}
	\label{fig:div2k4d}
\end{figure*}

In order to pave the way for the evaluation of blind SISR methods, we establish two datasets, including the synthetic DIV2K4D dataset which contains four subdatasets with a total of 400 images generated from the 100 DIV2K validation images with four different degradation types and the real RealSRSet which consists of 20 real images either downloaded from the internet or directly chosen from existing testing datasets~\cite{MartinFTM01,matsui2017sketch,zhang2018ffdnet,ignatov2017dslr}. Specifically, the four degradation types for DIV2K4D including 1) type I: the commonly-used bicubic degradation; 2) type II: anisotropic Gaussian blur with nearest downsampling by a scale factor of 4;  3) type III: anisotropic Gaussian blur with nearest downsampling by a scale factor of 2 and subsequent bicubic downsampling by another scale factor of 2 and final JPEG compression with quality factors uniformly sampled from $[41, 90]$; and 4) type IV: our proposed degradation model. Note that the subdataset with degradation type II and the downsampled images by a scale factor of 2 for subdataset with degradation type III are directly borrowed from the DIV2KRK dataset~\cite{bell2019blind}. Some example images from the two datasets are shown in Fig.~\ref{fig:dataset}, from which we can see the LR images are corrupted by diverse blur and noise degradations.
We argue that a general-purpose blind super-resolver should achieve a good overall performance on the two datasets.

\subsection{Compared Methods}
We compare the proposed BSRNet and BSRGAN with RRDB~\cite{wang2018esrgan}, IKC~\cite{gu2019blind}, ESRGAN~\cite{wang2018esrgan}, FSSR-DPED~\cite{fritsche2019frequency}, FSSR-JPEG~\cite{fritsche2019frequency}, RealSR-DPED~\cite{ji2020real} and RealSR-JPEG~\cite{ji2020real}. Specifically, RRDB and ESRGAN are trained on bicubic degradation; IKC is a blind model trained with different isotropic Gaussian kernels; FSSR-DPED and RealSR-DPED are trained to maximize the performance on the blurry and noisy DPED dataset; FSSR-JPEG is trained for JPEG image super-resolution; RealSR-JPEG is a recently released and unpublished model on github.
Note that since our novelty lies in the degradation model, and RRDB, ESRGAN, FSSR-DPED, FSSR-JPEG, RealSR-DPED and RealSR-JPEG use the same network architecture as ours, we thus did not re-train other models for comparison.

\begin{figure*}[!tbp]\footnotesize
\hspace{-0.20cm}
\begin{tabular}{c@{\extracolsep{0em}}c@{\extracolsep{0.05em}}c@{\extracolsep{0.05em}}c@{\extracolsep{0.05em}}c@{\extracolsep{0.05em}}@{\extracolsep{0.05em}}c@{\extracolsep{0.05em}}c}
        \includegraphics[width=0.16\textwidth]{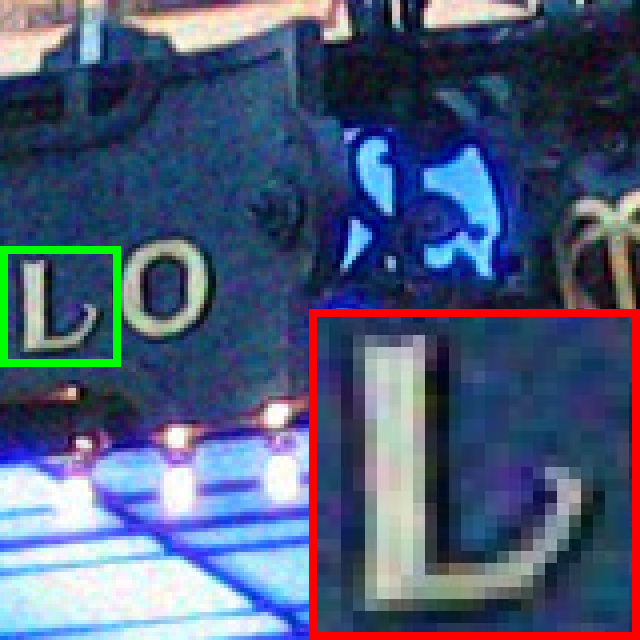}~
		&\includegraphics[width=0.16\textwidth]{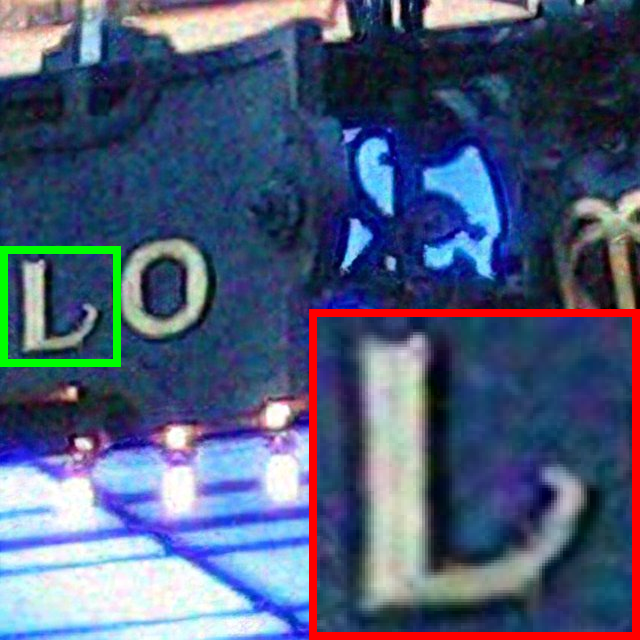}~
		&\includegraphics[width=0.16\textwidth]{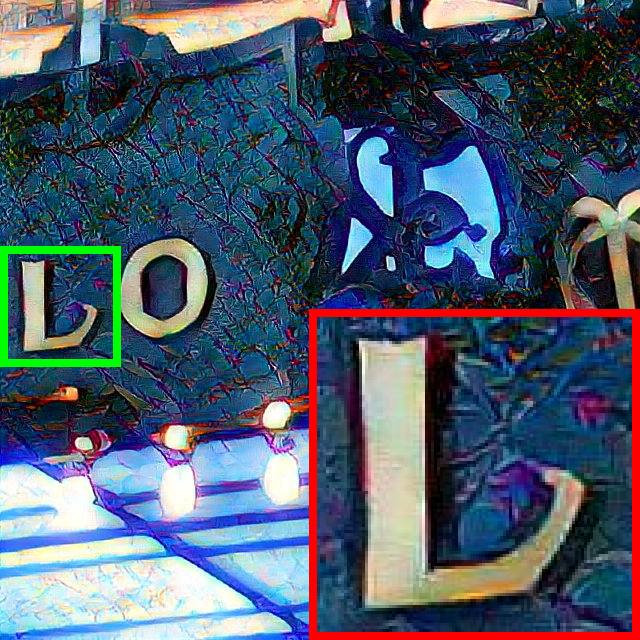}~
        &\includegraphics[width=0.16\textwidth]{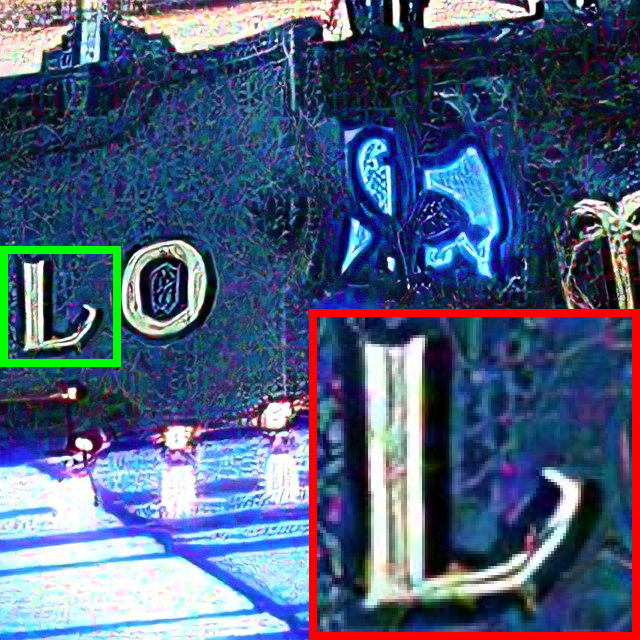}~
		&\includegraphics[width=0.16\textwidth]{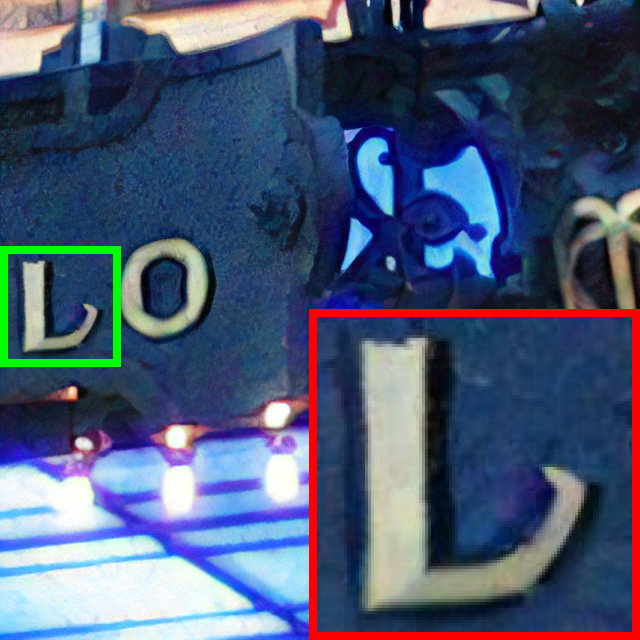}~
		&\includegraphics[width=0.16\textwidth]{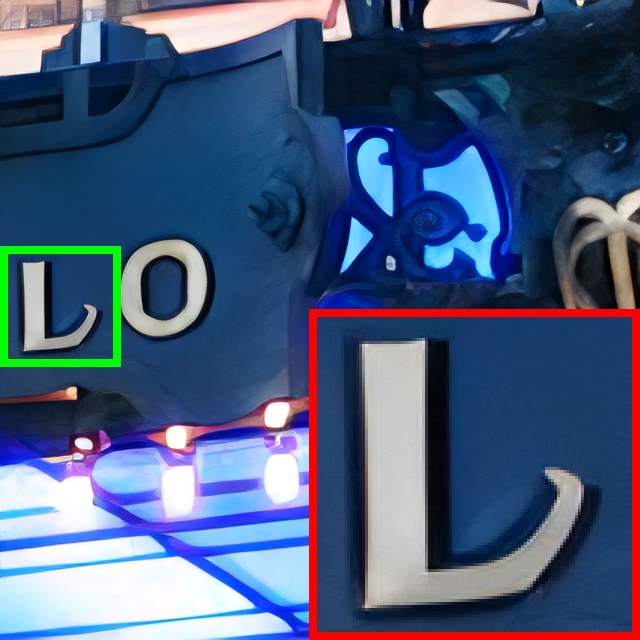}\\
		
NIQE$\downarrow$/NRQM$\uparrow$/PI$\downarrow$  &  4.47/3.15/5.65 & 4.19/\textcolor{red}{7.08}/\textcolor{blue}{3.55}  & \textcolor{red}{3.12}/\textcolor{blue}{6.81}/\textcolor{red}{3.15} & \textcolor{blue}{3.89}/4.39/4.75  & 4.52/5.79/4.36  \\

        \includegraphics[width=0.16\textwidth]{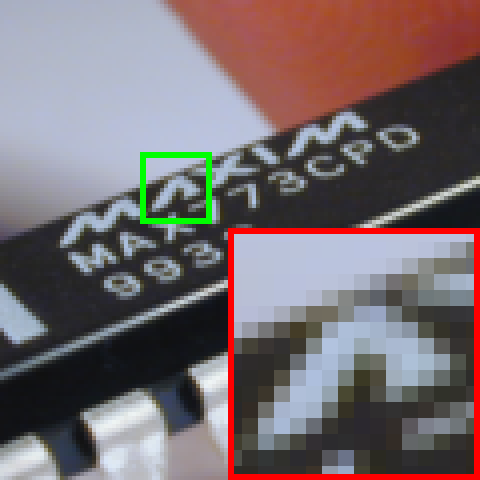}~
		&\includegraphics[width=0.16\textwidth]{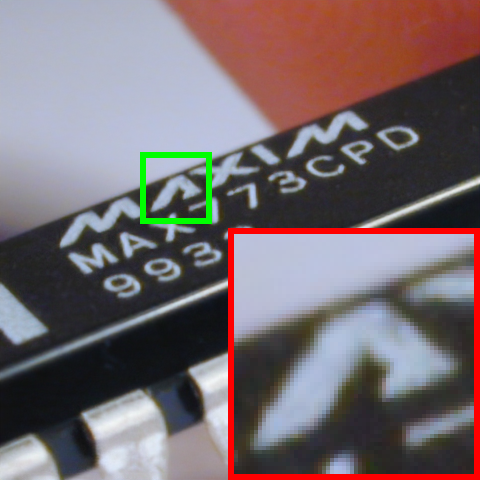}~
		&\includegraphics[width=0.16\textwidth]{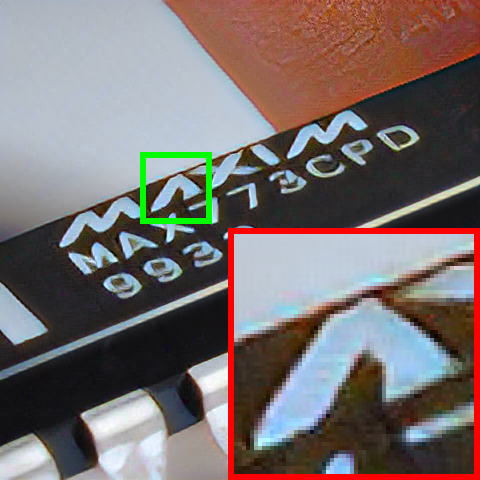}~
        &\includegraphics[width=0.16\textwidth]{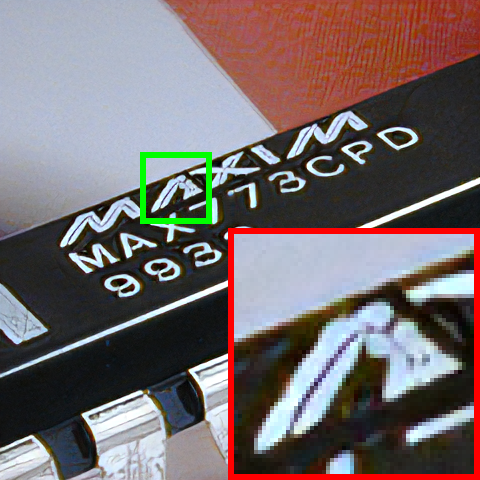}~
		&\includegraphics[width=0.16\textwidth]{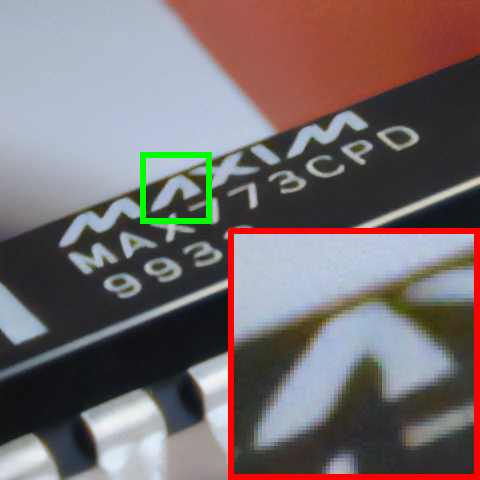}~
		&\includegraphics[width=0.16\textwidth]{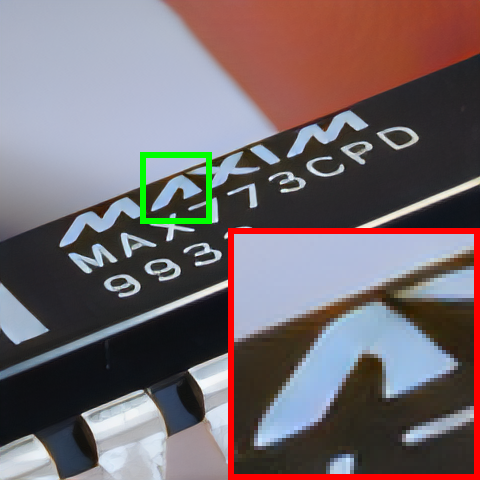}\\
NIQE$\downarrow$/NRQM$\uparrow$/PI$\downarrow$  & 5.85/4.66/5.59  &   \textcolor{red}{4.16}/\textcolor{red}{7.98}/\textcolor{red}{3.09} & \textcolor{blue}{4.64}/6.56/4.04  &  6.95/4.32/6.31 & 5.07/\textcolor{blue}{7.44}/\textcolor{blue}{3.82}  \\
        \includegraphics[width=0.16\textwidth]{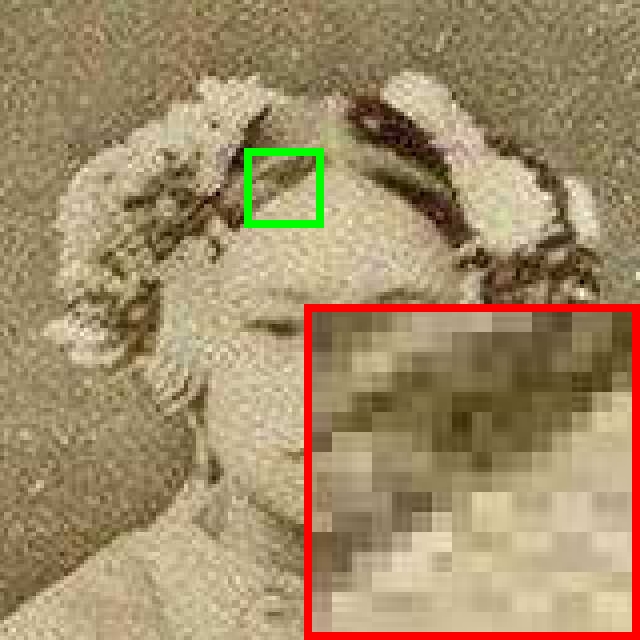}~
		&\includegraphics[width=0.16\textwidth]{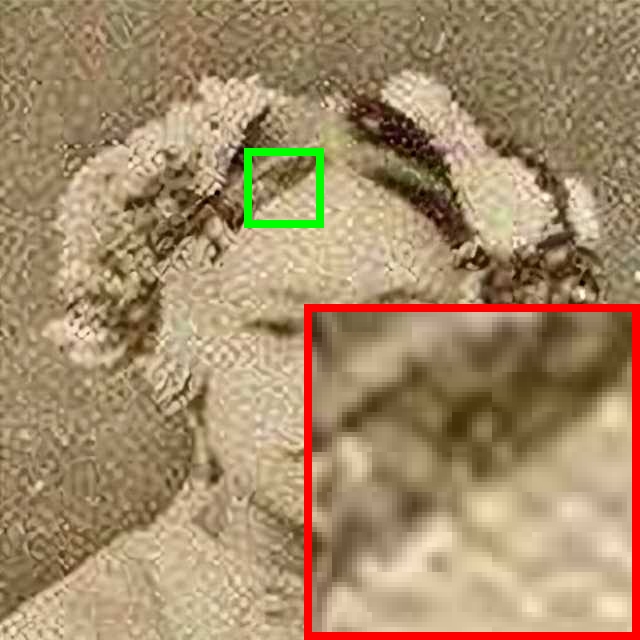}~
		&\includegraphics[width=0.16\textwidth]{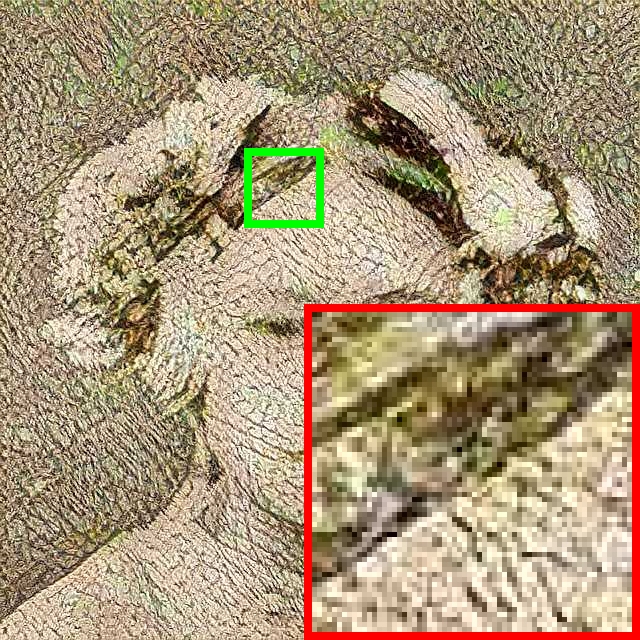}~
        &\includegraphics[width=0.16\textwidth]{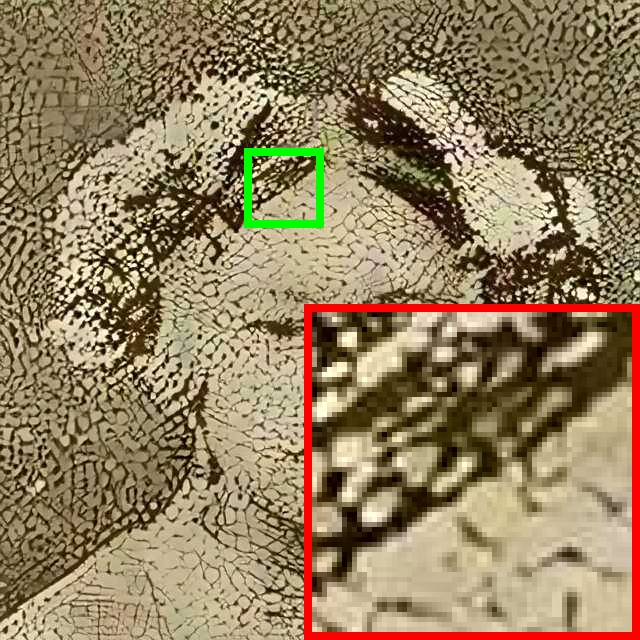}~
		&\includegraphics[width=0.16\textwidth]{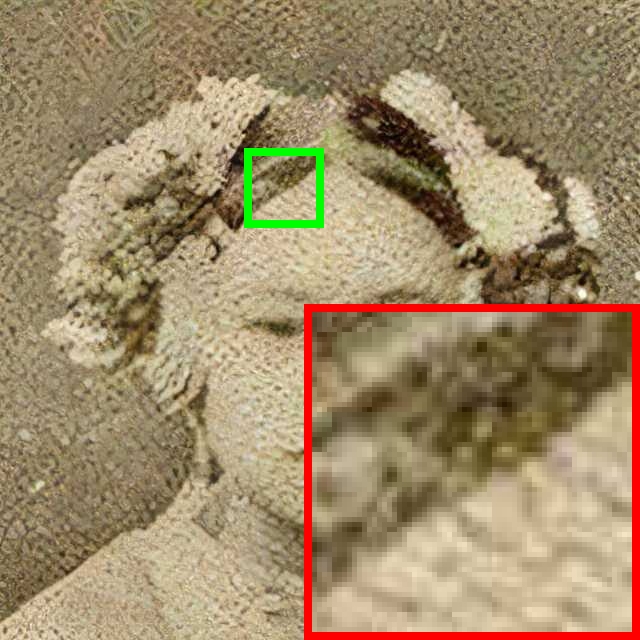}~
		&\includegraphics[width=0.16\textwidth]{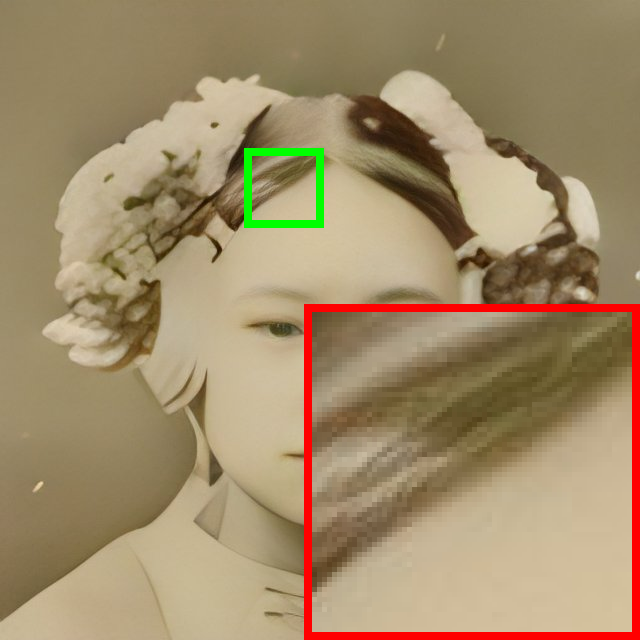}\\
NIQE$\downarrow$/NRQM$\uparrow$/PI$\downarrow$  & 7.10/3.92/6.59  &  \textcolor{blue}{5.31}/6.26/\textcolor{blue}{4.52} & 6.39/\textcolor{blue}{6.83}/4.78 &  \textcolor{red}{4.45}/\textcolor{red}{7.14}/\textcolor{red}{3.65} & 5.83/5.99/4.92 \\
(a) LR ($\times$4) &(b) ESRGAN~\cite{wang2018esrgan} &(c) FSSR-JPEG~\cite{fritsche2019frequency}   &(d) RealSR-DPED~\cite{ji2020real} & (e) RealSR-JPEG~\cite{ji2020real} &(f) \textbf{BSRGAN (Ours)} \\
	\end{tabular}
    \vspace{0.02cm}
	\caption{Results of different methods on super-resolving real images from RealSRSet with scale factor 4. The LR images from top to bottom in each row are ``\emph{Building}'', ``\emph{Chip}'', and ``\emph{Oldphoto2}'', respectively. Please zoom in for better view.}
	\label{fig:realsrset}
	\vspace{-0.2cm}
\end{figure*}

\subsection{Experiments on the DIV2K4D Dataset}

The PSNR and LPIPS (learned perceptual image patch similarity) results of different methods on the DIV2K4D datasets are shown in Table~\ref{table:div2k4d}. Note that LPIPS is used to measure the perceptual quality, and a lower LPIPS value means the super-resolved image is more perceptually similar to the ground-truth. We draw several conclusions from Table~\ref{table:div2k4d}. Firstly, as expected, RRDB and ESRGAN perform well for bicubic degradation but do not perform well on non-bicubic degradation as they are trained with the simplified bicubic degradation. It is worth noting that, even trained with GAN, ESRGAN can slightly improve the LPIPS values over RRDB on degradation types II-IV. Secondly, FSSR-DPED, FSSR-JPEG, RealSR-DPED and RealSR-JPEG outperform RRDB and ESRGAN in terms of LPIPS since they consider a more practical degradation. Thirdly, for degradation type II, IKC obtains promising PSNR results while RealSR-DPED achieves the best LPIPS result as they are trained on a similar degradation. For degradation types III and IV, they suffer a severe performance drop.
Fourthly, our proposed BSRNet achieves the best overall PSNR results, while BSRGAN yields the best overall LPIPS results.

Fig.~\ref{fig:div2k4d} shows the results of different methods on super-resolving an LR image from the DIV2K4D dataset. It can be seen that IKC and RealSR-JPEG fail to remove the noise and to recover sharp edges. On the other hand, FSSR-JPEG can produce sharp images but also introduces some artifacts. In comparison, our BSRNet and BSRGAN produce better visual results than the other methods.

\subsection{Experiments on the RealSRSet Dataset}

Since the ground-truth for the RealSRSet dataset is not available, we adopt the non-reference image quality assessment (IQA) metrics including NIQE~\cite{mittal2012making}, NRQM~\cite{ma2017learning} and PI~\cite{blau20182018} for quantitative evaluation. As one can see from Table~\ref{table:realsrset}, BSRGAN fails to show promising results. Yet, as shown in Fig.~\ref{fig:realsrset}, BSRNet produces much better visual results than the other methods. For example, BSRGAN can remove the unknown processed camera sensor noise for ``\emph{Building}'' and unknown complex noise for ``\emph{Oldphoto2}'', while also producing sharp edges and fine details. In contrast, FSSR-JPEG, RealSR-DPED and RealSR-JPEG produce some high-frequency artifacts but have better quantitative results than BSRNet.
Such inconsistencies indicate that these no-reference IQA metrics do not always match perceptual visual quality~\cite{lugmayr2020ntire} and the IQA metric could be updated with new SISR methods~\cite{gu2020pipal}.
We further argue that the IQA metric for SISR should also be updated with new image degradation types, which we leave for future work.
We note that our BSRGAN tends to produce `bubble' artifacts in texture region, which may be solved by new loss function or more training data with diverse textures.

\begin{table}[!tbp]\scriptsize
\vspace{0.04cm}
\caption{The no-reference NIQE~\cite{mittal2012making}, NRQM~\cite{ma2017learning} and PI~\cite{blau20182018} results of different methods on the RealSRSet dataset. The best and second best results are highlighted in red and blue, respectively. Note that all the methods use the same network architecture.}
\center
\begin{tabular}{l|p{0.75cm}<{\centering}|p{.73cm}<{\centering}|p{.66cm}<{\centering}|p{.73cm}<{\centering}|p{.66cm}<{\centering}|p{0.8cm}<{\centering}}
\hline
\footnotesize{\multirow{2}{*}{\;Metric}}&\multirow{2}{*}{\hspace{-0.08cm}ESRGAN}& FSSR & FSSR & RealSR & RealSR &\textbf{BSRGAN}\\
& & \scriptsize -DPED & \scriptsize -JPEG & \scriptsize -DPED & \scriptsize -JPEG & \textbf{(Ours)}\\ \hline \hline
\,NIQE$\downarrow$ & \footnotesize 4.95 &\footnotesize 4.86 & \footnotesize \textcolor{blue}{4.04}  &\footnotesize  4.58  & \footnotesize \textcolor{red}{3.99} & \footnotesize   5.60\\
NRQM$\uparrow$ &\footnotesize  6.02 &\footnotesize  6.28 & \footnotesize \textcolor{red}{6.88} &\footnotesize  \textcolor{blue}{6.59}  &\footnotesize  6.23 &\footnotesize  6.17  \\
\;\;PI$\downarrow$ &\footnotesize  4.47 &\footnotesize  4.29 &\footnotesize  \textcolor{red}{3.58} &\footnotesize  \textcolor{blue}{3.99}  &\footnotesize  4.29 &\footnotesize  4.72  \\
\hline
\end{tabular}
\vspace{-0.2cm}
\label{table:realsrset}
\end{table}

\section{Conclusions}
\label{sec:conclusion}
In this paper, we have designed a new degradation model to train a deep blind super-resolution model. Specifically, by making each of the degradation factors, \ie blur, downsampling and noise, more intricate and practical, and also by introducing a random shuffle strategy, the new degradation model can cover a wide range of degradations found in real-world scenarios. Based on the synthetic data generated by the new degradation model, we have trained a deep blind model for general image super-resolution. Experiments on synthetic and real image datasets have shown that the deep blind model performs favorably on images corrupted by diverse degradations.
We believe that existing deep super-resolution networks can benefit from our new degradation model to enhance their usefulness in practice. As a result, this work provides a way towards solving blind super-resolution for real applications.

\noindent\textbf{Acknowledgments:}
This work was partly supported by the ETH Z\"urich Fund (OK), a Huawei Technologies Oy (Finland) project, and an Amazon AWS grant.


{\small
\bibliographystyle{ieee_fullname}
\bibliography{egbib}

\begin{thebibliography}{10}\itemsep=-1pt

\bibitem{adobe2019}
Adobe.
\newblock Digital negative specification.
\newblock 2019.
\newblock Version 1.5.00.

\bibitem{agustsson2017ntire}
Eirikur Agustsson and Radu Timofte.
\newblock Ntire 2017 challenge on single image super-resolution: Dataset and
  study.
\newblock In {\em CVPR Workshops}, volume~3, pages 126--135, July 2017.

\bibitem{bell2019blind}
Sefi Bell-Kligler, Assaf Shocher, and Michal Irani.
\newblock Blind super-resolution kernel estimation using an internal-gan.
\newblock In {\em NeurIPS}, pages 284--293, 2019.

\bibitem{blau20182018}
Yochai Blau, Roey Mechrez, Radu Timofte, Tomer Michaeli, and Lihi Zelnik-Manor.
\newblock The 2018 {PIRM} challenge on perceptual image super-resolution.
\newblock In {\em ECCV Workshops}, 2018.

\bibitem{blau2018perception}
Yochai Blau and Tomer Michaeli.
\newblock The perception-distortion tradeoff.
\newblock In {\em CVPR}, pages 6228--6237, 2018.

\bibitem{brooks2019unprocessing}
Tim Brooks, Ben Mildenhall, Tianfan Xue, Jiawen Chen, Dillon Sharlet, and
  Jonathan~T Barron.
\newblock Unprocessing images for learned raw denoising.
\newblock In {\em CVPR}, pages 11036--11045, 2019.

\bibitem{cai2019toward}
Jianrui Cai, Hui Zeng, Hongwei Yong, Zisheng Cao, and Lei Zhang.
\newblock Toward real-world single image super-resolution: A new benchmark and
  a new model.
\newblock In {\em ICCV}, pages 3086--3095, 2019.

\bibitem{cornillere2019blind}
Victor Cornillere, Abdelaziz Djelouah, Wang Yifan, Olga Sorkine-Hornung, and
  Christopher Schroers.
\newblock Blind image super-resolution with spatially variant degradations.
\newblock {\em ACM TOG}, 38(6):1--13, 2019.

\bibitem{cubuk2020randaugment}
Ekin~D Cubuk, Barret Zoph, Jonathon Shlens, and Quoc~V Le.
\newblock Randaugment: Practical automated data augmentation with a reduced
  search space.
\newblock In {\em CVPR Workshops}, pages 702--703, 2020.

\bibitem{dong2014learning}
Chao Dong, Chen~Change Loy, Kaiming He, and Xiaoou Tang.
\newblock Learning a deep convolutional network for image super-resolution.
\newblock In {\em ECCV}, pages 184--199, 2014.

\bibitem{dong2013nonlocally}
Weisheng Dong, Lei Zhang, Guangming Shi, and Xin Li.
\newblock Nonlocally centralized sparse representation for image restoration.
\newblock {\em IEEE TIP}, 22(4):1620--1630, 2013.

\bibitem{efrat2013accurate}
Netalee Efrat, Daniel Glasner, Alexander Apartsin, Boaz Nadler, and Anat Levin.
\newblock Accurate blur models vs. image priors in single image
  super-resolution.
\newblock In {\em ICCV}, pages 2832--2839, 2013.

\bibitem{fritsche2019frequency}
Manuel Fritsche, Shuhang Gu, and Radu Timofte.
\newblock Frequency separation for real-world super-resolution.
\newblock In {\em ICCV Workshop}, pages 3599--3608, 2019.

\bibitem{grossberg2003space}
Michael~D Grossberg and Shree~K Nayar.
\newblock What is the space of camera response functions?
\newblock In {\em CVPR}, pages II--602, 2003.

\bibitem{gu2020pipal}
Jinjin Gu, Haoming Cai, Haoyu Chen, Xiaoxing Ye, Jimmy Ren, and Chao Dong.
\newblock Pipal: a large-scale image quality assessment dataset for perceptual
  image restoration.
\newblock {\em ECCV}, 2020.

\bibitem{gu2019blind}
Jinjin Gu, Hannan Lu, Wangmeng Zuo, and Chao Dong.
\newblock Blind super-resolution with iterative kernel correction.
\newblock In {\em CVPR}, pages 1604--1613, 2019.

\bibitem{hui2019lightweight}
Zheng Hui, Xinbo Gao, Yunchu Yang, and Xiumei Wang.
\newblock Lightweight image super-resolution with information
  multi-distillation network.
\newblock In {\em ICME}, pages 2024--2032, 2019.

\bibitem{ignatov2017dslr}
Andrey Ignatov, Nikolay Kobyshev, Radu Timofte, Kenneth Vanhoey, and Luc
  Van~Gool.
\newblock {DSLR}-quality photos on mobile devices with deep convolutional
  networks.
\newblock In {\em ICCV}, pages 3277--3285, 2017.

\bibitem{isola2017image}
Phillip Isola, Jun-Yan Zhu, Tinghui Zhou, and Alexei~A Efros.
\newblock Image-to-image translation with conditional adversarial networks.
\newblock In {\em CVPR}, pages 1125--1134, 2017.

\bibitem{ji2020real}
Xiaozhong Ji, Yun Cao, Ying Tai, Chengjie Wang, Jilin Li, and Feiyue Huang.
\newblock Real-world super-resolution via kernel estimation and noise
  injection.
\newblock In {\em CVPR Workshops}, pages 466--467, 2020.

\bibitem{jiang2021towards}
Jiaxi Jiang, Kai Zhang, and Radu Timofte.
\newblock Towards flexible blind {JPEG} artifacts removal.
\newblock In {\em ICCV}, 2021.

\bibitem{karras2019style}
Tero Karras, Samuli Laine, and Timo Aila.
\newblock A style-based generator architecture for generative adversarial
  networks.
\newblock In {\em CVPR}, pages 4401--4410, 2019.

\bibitem{lai2017deep}
Wei-Sheng Lai, Jia-Bin Huang, Narendra Ahuja, and Ming-Hsuan Yang.
\newblock Deep laplacian pyramid networks for fast and accurate
  super-resolution.
\newblock In {\em CVPR}, pages 624--632, July 2017.

\bibitem{ledig2016photo}
Christian Ledig, Lucas Theis, Ferenc Husz{\'a}r, Jose Caballero, Andrew
  Cunningham, Alejandro Acosta, Andrew Aitken, Alykhan Tejani, Johannes Totz,
  Zehan Wang, et~al.
\newblock Photo-realistic single image super-resolution using a generative
  adversarial network.
\newblock In {\em CVPR}, pages 4681--4690, July 2017.

\bibitem{liang2021hierarchical}
Jingyun Liang, Andreas Lugmayr, Kai Zhang, Martin Danelljan, Luc Van~Gool, and
  Radu Timofte.
\newblock Hierarchical conditional flow: A unified framework for image
  super-resolution and image rescaling.
\newblock In {\em ICCV}, 2021.

\bibitem{liang2021flow}
Jingyun Liang, Kai Zhang, Shuhang Gu, Luc Van~Gool, and Radu Timofte.
\newblock Flow-based kernel prior with application to blind super-resolution.
\newblock In {\em CVPR}, pages 10601--10610, 2021.

\bibitem{lim2017enhanced}
Bee Lim, Sanghyun Son, Heewon Kim, Seungjun Nah, and Kyoung~Mu Lee.
\newblock Enhanced deep residual networks for single image super-resolution.
\newblock In {\em CVPR Workshops}, pages 136--144, July 2017.

\bibitem{liu2013bayesian}
Ce Liu and Deqing Sun.
\newblock On bayesian adaptive video super resolution.
\newblock {\em IEEE Transactions on Pattern Analysis and Machine Intelligence},
  36(2):346--360, 2013.

\bibitem{lugmayr2019unsupervised}
Andreas {Lugmayr}, Martin {Danelljan}, and Radu {Timofte}.
\newblock Unsupervised learning for real-world super-resolution.
\newblock In {\em ICCV Workshop}, pages 3408--3416, 2019.

\bibitem{lugmayr2020ntire}
Andreas Lugmayr, Martin Danelljan, and Radu Timofte.
\newblock Ntire 2020 challenge on real-world image super-resolution: Methods
  and results.
\newblock In {\em CVPR Workshops}, pages 494--495, 2020.

\bibitem{luo2020unfolding}
Zhengxiong Luo, Yan Huang, Shang Li, Liang Wang, and Tieniu Tan.
\newblock Unfolding the alternating optimization for blind super resolution.
\newblock {\em NeurIPS}, 33, 2020.

\bibitem{ma2017learning}
Chao Ma, Chih-Yuan Yang, Xiaokang Yang, and Ming-Hsuan Yang.
\newblock Learning a no-reference quality metric for single-image
  super-resolution.
\newblock {\em CVIU}, 158:1--16, 2017.

\bibitem{ma2016gmad}
Kede Ma, Zhengfang Duanmu, Qingbo Wu, Zhou Wang, Hongwei Yong, Hongliang Li,
  and Lei Zhang.
\newblock Waterloo exploration database: New challenges for image quality
  assessment models.
\newblock {\em IEEE TIP}, 26(2):1004--1016, 2017.

\bibitem{malvar2004high}
Henrique~S Malvar, Li-wei He, and Ross Cutler.
\newblock High-quality linear interpolation for demosaicing of bayer-patterned
  color images.
\newblock In {\em IEEE International Conference on Acoustics, Speech, and
  Signal Processing}, volume~3, pages iii--485, 2004.

\bibitem{MartinFTM01}
D. Martin, C. Fowlkes, D. Tal, and J. Malik.
\newblock A database of human segmented natural images and its application to
  evaluating segmentation algorithms and measuring ecological statistics.
\newblock In {\em ICCV}, volume~2, pages 416--423, 2001.

\bibitem{matsui2017sketch}
Yusuke Matsui, Kota Ito, Yuji Aramaki, Azuma Fujimoto, Toru Ogawa, Toshihiko
  Yamasaki, and Kiyoharu Aizawa.
\newblock Sketch-based manga retrieval using manga109 dataset.
\newblock {\em Multimedia Tools and Applications}, 76(20):21811--21838, 2017.

\bibitem{michaeli2013nonparametric}
Tomer Michaeli and Michal Irani.
\newblock Nonparametric blind super-resolution.
\newblock In {\em ICCV}, pages 945--952, 2013.

\bibitem{mittal2012making}
Anish Mittal, Rajiv Soundararajan, and Alan~C Bovik.
\newblock Making a ``completely blind'' image quality analyzer.
\newblock {\em IEEE SPL}, 20(3):209--212, 2012.

\bibitem{nam2016holistic}
Seonghyeon Nam, Youngbae Hwang, Yasuyuki Matsushita, and Seon~Joo Kim.
\newblock A holistic approach to cross-channel image noise modeling and its
  application to image denoising.
\newblock In {\em CVPR}, pages 1683--1691, 2016.

\bibitem{park2013gaussian}
Sangwoo Park, Erchin Serpedin, and Khalid Qaraqe.
\newblock Gaussian assumption: The least favorable but the most useful [lecture
  notes].
\newblock {\em IEEE SPM}, 30(3):183--186, 2013.

\bibitem{peleg2014statistical}
Tomer Peleg and Michael Elad.
\newblock A statistical prediction model based on sparse representations for
  single image super-resolution.
\newblock {\em IEEE TIP}, 23(6):2569--2582, 2014.

\bibitem{plotz2017benchmarking}
Tobias Plotz and Stefan Roth.
\newblock Benchmarking denoising algorithms with real photographs.
\newblock In {\em CVPR}, pages 1586--1595, 2017.

\bibitem{riegler2015conditioned}
Gernot Riegler, Samuel Schulter, Matthias Ruther, and Horst Bischof.
\newblock Conditioned regression models for non-blind single image
  super-resolution.
\newblock In {\em ICCV}, pages 522--530, 2015.

\bibitem{sajjadi2017enhancenet}
Mehdi~SM Sajjadi, Bernhard Sch{\"o}lkopf, and Michael Hirsch.
\newblock Enhancenet: Single image super-resolution through automated texture
  synthesis.
\newblock In {\em ICCV}, pages 4501--4510, 2017.

\bibitem{shocher2018zero}
Assaf Shocher, Nadav Cohen, and Michal Irani.
\newblock ``zero-shot'' super-resolution using deep internal learning.
\newblock In {\em ICCV}, pages 3118--3126, 2018.

\bibitem{timofte2017ntire}
Radu Timofte, Eirikur Agustsson, Luc Van~Gool, Ming-Hsuan Yang, and Lei Zhang.
\newblock Ntire 2017 challenge on single image super-resolution: Methods and
  results.
\newblock In {\em CVPR Workshops}, pages 114--125, 2017.

\bibitem{timofte2014a+}
Radu Timofte, Vincent De~Smet, and Luc Van~Gool.
\newblock A+: Adjusted anchored neighborhood regression for fast
  super-resolution.
\newblock In {\em ACCV}, pages 111--126, 2014.

\bibitem{wang2021unsupervised}
Longguang Wang, Yingqian Wang, Xiaoyu Dong, Qingyu Xu, Jungang Yang, Wei An,
  and Yulan Guo.
\newblock Unsupervised degradation representation learning for blind
  super-resolution.
\newblock In {\em CVPR}, pages 10581--10590, 2021.

\bibitem{wang2018esrgan}
Xintao Wang, Ke Yu, Shixiang Wu, Jinjin Gu, Yihao Liu, Chao Dong, Yu Qiao, and
  Chen~Change Loy.
\newblock {ESRGAN}: Enhanced super-resolution generative adversarial networks.
\newblock In {\em ECCV Workshops}, 2018.

\bibitem{wei2020aim}
Pengxu Wei, Hannan Lu, Radu Timofte, Liang Lin, Wangmeng Zuo, et~al.
\newblock Aim 2020 challenge on real image super-resolution: Methods and
  results.
\newblock In {\em ECCV Workshops}, 2020.

\bibitem{yuan2018unsupervised}
Yuan Yuan, Siyuan Liu, Jiawei Zhang, Yongbing Zhang, Chao Dong, and Liang Lin.
\newblock Unsupervised image super-resolution using cycle-in-cycle generative
  adversarial networks.
\newblock In {\em CVPR Workshops}, pages 701--710, 2018.

\bibitem{zhang2020deep}
Kai Zhang, Luc~Van Gool, and Radu Timofte.
\newblock Deep unfolding network for image super-resolution.
\newblock In {\em CVPR}, pages 3217--3226, 2020.

\bibitem{zhang2021plug}
Kai Zhang, Yawei Li, Wangmeng Zuo, Lei Zhang, Luc Van~Gool, and Radu Timofte.
\newblock Plug-and-play image restoration with deep denoiser prior.
\newblock {\em IEEE TPAMI}, 2021.

\bibitem{zhang2015revisiting}
Kai Zhang, Xiaoyu Zhou, Hongzhi Zhang, and Wangmeng Zuo.
\newblock Revisiting single image super-resolution under internet environment:
  blur kernels and reconstruction algorithms.
\newblock In {\em Pacific Rim Conference on Multimedia}, pages 677--687, 2015.

\bibitem{zhang2017beyond}
Kai Zhang, Wangmeng Zuo, Yunjin Chen, Deyu Meng, and Lei Zhang.
\newblock Beyond a gaussian denoiser: Residual learning of deep {CNN} for image
  denoising.
\newblock {\em IEEE TIP}, pages 3142--3155, 2017.

\bibitem{zhang2017learning}
Kai Zhang, Wangmeng Zuo, Shuhang Gu, and Lei Zhang.
\newblock Learning deep {CNN} denoiser prior for image restoration.
\newblock In {\em CVPR}, pages 3929--3938, July 2017.

\bibitem{zhang2018ffdnet}
Kai Zhang, Wangmeng Zuo, and Lei Zhang.
\newblock {FFDNet}: Toward a fast and flexible solution for {CNN}-based image
  denoising.
\newblock {\em IEEE TIP}, 27(9):4608--4622, 2018.

\bibitem{zhang2018learning}
Kai Zhang, Wangmeng Zuo, and Lei Zhang.
\newblock Learning a single convolutional super-resolution network for multiple
  degradations.
\newblock In {\em CVPR}, pages 3262--3271, 2018.

\bibitem{zhang2019deep}
Kai Zhang, Wangmeng Zuo, and Lei Zhang.
\newblock Deep plug-and-play super-resolution for arbitrary blur kernels.
\newblock In {\em CVPR}, pages 1671--1681, 2019.

\bibitem{zhang2018image}
Yulun Zhang, Kunpeng Li, Kai Li, Lichen Wang, Bineng Zhong, and Yun Fu.
\newblock Image super-resolution using very deep residual channel attention
  networks.
\newblock In {\em ECCV}, pages 286--301, 2018.

\bibitem{zhang2018residual}
Yulun Zhang, Yapeng Tian, Yu Kong, Bineng Zhong, and Yun Fu.
\newblock Residual dense network for image super-resolution.
\newblock In {\em ICCV}, pages 2472--2481, 2018.

\end{thebibliography}
}

\end{document}